\documentclass[preprint]{aastex}

\shorttitle{Multi-wavelength variability monitoring}
\shortauthors{Biller et al.}

\begin{document}


\title{Weather on the Nearest Brown Dwarfs: \\ Resolved 
Simultaneous Multi-Wavelength Variability Monitoring \\ of WISE J104915.57-531906.1AB}


\author{Beth A. Biller\altaffilmark{1,2}, Ian J.M.
  Crossfield\altaffilmark{1}, Luigi Mancini\altaffilmark{1}, 
  Simona Ciceri\altaffilmark{1}, John Southworth\altaffilmark{3}, 
  Taisiya G. Kopytova\altaffilmark{1,4}, 
  Micka\"el Bonnefoy\altaffilmark{1}, Niall R. Deacon\altaffilmark{1}, 
  Joshua E. Schlieder\altaffilmark{1}, Esther Buenzli\altaffilmark{1}, 
  Wolfgang Brandner\altaffilmark{1}, France Allard\altaffilmark{5}, 
  Derek Homeier\altaffilmark{5}, Bernd Freytag\altaffilmark{6}, 
  Coryn A.L. Bailer-Jones\altaffilmark{1}, 
  Jochen Greiner\altaffilmark{7}, Thomas Henning\altaffilmark{1}, 
  Bertrand Goldman\altaffilmark{1}}

\altaffiltext{1}{Max-Planck-Institut f\"ur Astronomie, K\"onigstuhl
  17, 69117 Heidelberg, Germany}
\altaffiltext{2}{Institute for Astronomy, University of Edinburgh, Blackford Hill View, Edinburgh EH9 3HJ, UK}
\altaffiltext{3}{Astrophysics Group, Keele University, Staffordshire, ST5 5BG, UK}
\altaffiltext{4}{International Max-Planck Research
School for Astronomy and Cosmic Physics at the University of Heidelberg, IMPRS-HD,
Germany}
\altaffiltext{5}{Centre de Recherche Astrophysique de Lyon, {\'E}cole Normale Sup{\'e}rieure de Lyon, 46 all{\'e}e d'Italie, 69364 Lyon cedex 07, France}
\altaffiltext{6}{Department of Physics and Astronomy, Uppsala Universitet,
Regementsv{\"a}gen 1, Box 515, 75120 Uppsala, Sweden}
\altaffiltext{7}{Max-Planck Institute for extraterrestrische Physik, 85748 Garching,
  Giessenbachstr., Germany}







\begin{abstract}
We present two epochs of MPG/ESO 2.2m GROND simultaneous 6-band
($r'i'z'JHK$) photometric 
monitoring of the closest known L/T transition brown dwarf binary 
WISE J104915.57-531906.1AB.  We report here the first resolved 
variability monitoring of both the T0.5 and L7.5 components.   
We obtained 4 hours of focused observations on the night of UT
2013-04-22, as well as 4 hours of defocused (unresolved) observations 
on the night of UT 2013-04-16.  We note a number of robust trends in our light curves.
The $r'$ and $i'$ light curves appear to be anticorrelated 
with $z'$ and $H$ for the T0.5 component and in the 
unresolved lightcurve.  In the defocused dataset, $J$ appears correlated with $z'$ 
and $H$ and anticorrelated with $r'$ and $i'$, while in the focused dataset
we measure no variability for $J$ at the level of our photometric
precision, likely due to evolving weather phenomena.
In our focused T0.5 component lightcurve, the $K$ band lightcurve displays a
significant phase offset relative to both $H$ and $z'$.  We argue
that the measured phase offsets are correlated with atmospheric 
pressure probed at each band, as estimated from 1D
atmospheric models.  We also report low-amplitude
variability in $i'$ and $z'$ intrinsic to the L7.5 component. 
\end{abstract}





\section{Introduction}

Luhman (2013) recently reported the discovery of 
WISE J104915.57-531906.1AB (henceforth WISE
1049AB), a brown dwarf binary (1.5", 3 AU separation) with an 
L7.5 primary and a T0.5 secondary (Kniazev et al. 2013, Burgasser et
al. 2013).  WISE 1049AB are the closest and brightest brown dwarfs
in the sky (distance of 2.0$\pm$0.15 pc, combined $JHK$=10.7, 9.6,
8.8), enabling a level of detailed characterization impossible for fainter brown dwarfs.

The L/T transition spectral types of WISE 1049AB 
make them ideal targets to search for and characterize photometric
variability; at these effective temperatures the red dusty
clouds found in L dwarf atmospheres are expected to start breaking 
up (Marley et al. 2010).  Time-variability (brightness as a function 
of rotation angle) is a key diagnostic of cloud properties in brown dwarf
atmospheres, probing the spatial
distribution of condensates as the brown dwarf rotates. 
For instance, Artigau et al. (2009) observed correlated variability in
the $J$ and $K$ band (50 mmag amplitude in $J$, $\sim$2.4 hour period) for the 
L/T transition object (spectral type T2.5) SIMP J013656.57+093347.3.
This variability is best explained by 
a mixture of thick and thin patchy iron and silicate clouds covering the surface of this
object (Apai et al. 2013).   Quasi-periodic variability attributed to
clouds has already been observed for a number of L and T type field
brown dwarfs (Artigau et al. 2009, Radigan et al. 2012, Buenzli et al. 2012, Heinze
et al. 2013), with one object, 2MASS J21392676+0220226, 
displaying a peak-to-peak variability amplitude of 27\% (Radigan et al. 2012). 
Recent large-scale surveys of brown dwarf variability with Spitzer 
have revealed mid-IR variability in nearly two dozen L and T type brown
dwarfs (Heinze et al. 2013).  Buenzli et al. (2013, submitted) find that 
$\sim$30\% of the L5-T6 objects surveyed in their HST SNAP
survey show variability trends.  While variability is observed across
the full range of L and T spectral types, L/T transition brown dwarfs 
often possess higher amplitude variability than earlier L or later T
brown dwarfs (Radigan et al. in prep, Metchev et al. 2013).

Gillon et al. (2013) report a detection 
of 11$\%$ maximal peak-to-peak variability at 0.9 $\mu$m for WISE
1049AB, based on 12 nights
of photometric monitoring in an $I$+$z'$ filter at the TRAPPIST robotic
telescope (Jehin et al. 2011).  They attribute the measured
variability (with 4.87$\pm$0.01 hour period) to the cooler
T0.5 component, although the two components are barely resolved 
in their data.  While variability is clearly established for this
binary, resolved observations at multiple wavelengths are necessary
to characterize the variability and the cloud structure which presumably 
causes it.  

Here we present simultaneous multi-color, broad-band 
photometric monitoring of WISE 1049AB using GROND at the MPG/ESO 2.2 m.  
GROND is a unique 7-channel dual optical-IR camera
(Greiner et al. 2008), which images simultaneously in an optical camera
(four optical bands similar to Sloan $g'$$r'$$i'$$z'$, platescale 0.158$\arcsec$/pixel)
and an IR camera (three near-IR bands $JHK$, platescale 0.6$\arcsec$/pixel). 
Multi-band monitoring allows instantaneous
confirmation of variability across a wide spectral range and 
provides a rich dataset for further atmospheric characterization
of these objects.

\section{Observations and Data Reduction}

On UT 2013-04-16, 
we obtained 4 hours on sky of defocused observations (usable
coverage of 3.6 hours). 
Defocused observations are standard in transit observations
of exoplanets in order to avoid saturation and maximize photometric
precision (e.g. Southworth et al. 2009); because of initial concerns  
about resolving the binary in the near-IR with GROND's
0.6'' infrared pixels, we attempted a similar 
strategy for these very bright brown dwarfs, at the expense of resolved lightcurves.
On UT 2013-04-22, we obtained 4 hours on sky of
focused observations (usable coverage of 3.4 hours).  
In the focused data, counts in the two brown dwarfs  
were kept below saturation in all filters.  In both cases, no dither pattern
was used, in order to place the brown dwarfs reliably on the same
pixels and enable high-precision photometry.  
Optical observations were taken with a base exposure time 
of  100 s for the defocused observations and 114.9 s 
for the focused observations.  Infrared observations were taken with a base
exposure time of 2 s and with 10 coadded exposures
between readouts in both datasets.  
WISE 1049AB was detected in each individual frame in both datasets 
in $r'i'z'JHK$.  WISE1049AB was detected in $g'$ in the stacked
images in both datasets, but not at a suitable cadence for
variability monitoring.

\begin{figure}
\includegraphics[width=6.5in]{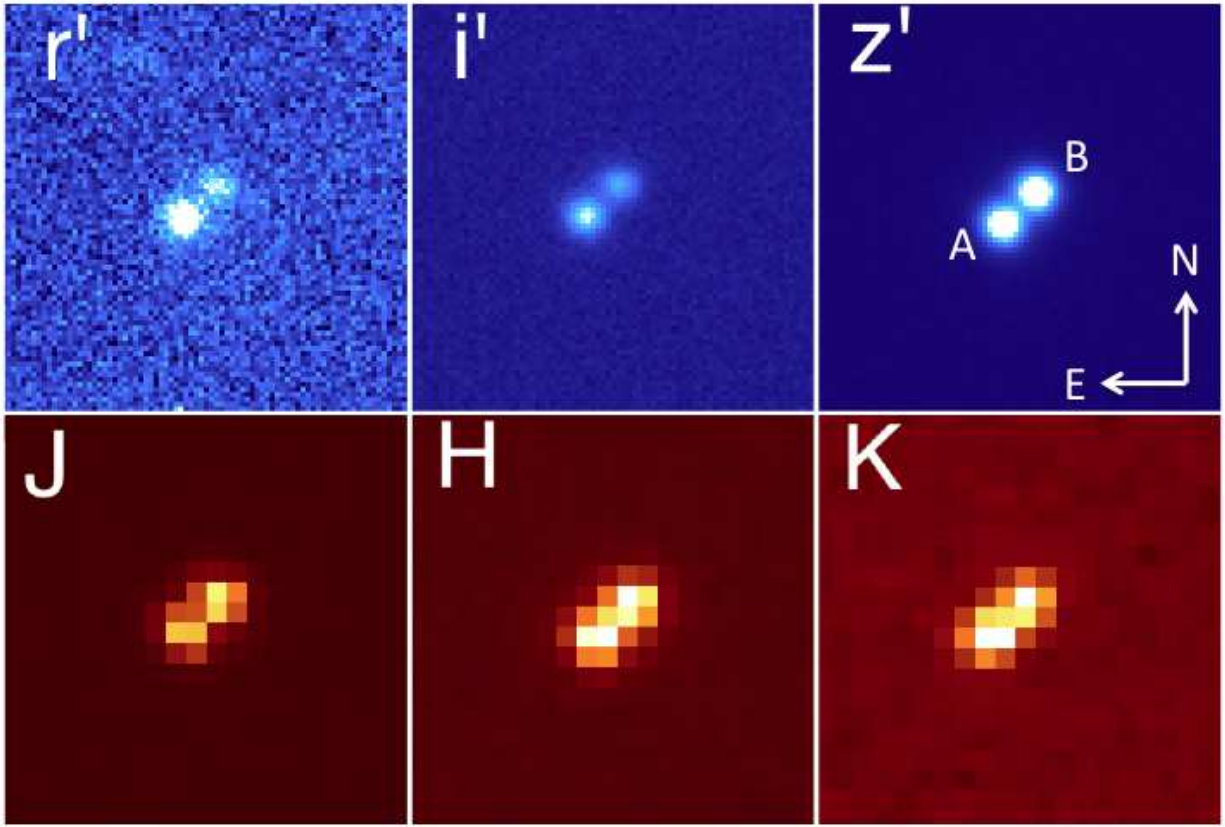}
\caption{Single frame optical and near-IR images in the 6 filters
  in which we detect WISE 1049AB ($r'$$i'$$z'$$JHK$).
The L7.5 component is southeast of the T0.5 component and the 
two components are separated by 1.5".  
\label{fig:images}
}
\end{figure}

Standard calibrations (bias subtraction, flat-fielding) were performed
on both datasets.  For the defocused dataset, we performed 
aperture photometry using the pipeline described in Southworth et al. (2009).
In the focused dataset, we resolve the binary in both the 
optical and near-IR data, although the PSF wings of the two 
components overlap in the near-IR (see Fig.~\ref{fig:images}).  We extracted 
resolved psf-fitting photometry using the StarFinder
IDL package (Diolaiti et al. 2000).  While the optical data 
was well-sampled, the near-IR data is undersampled 
with the 0.6'' platescale of the GROND IR detector.  Nonetheless,
with more than 20 resolved sources in the field,
we were able to generate an appropriate PSF and obtain 
reasonable subtractions from our PSF-fitting photometry.


\section{Light Curves}

\subsection{Unresolved combined-light light curves\label{sec:unresolved}}

\begin{figure}
\includegraphics[width=3.5in]{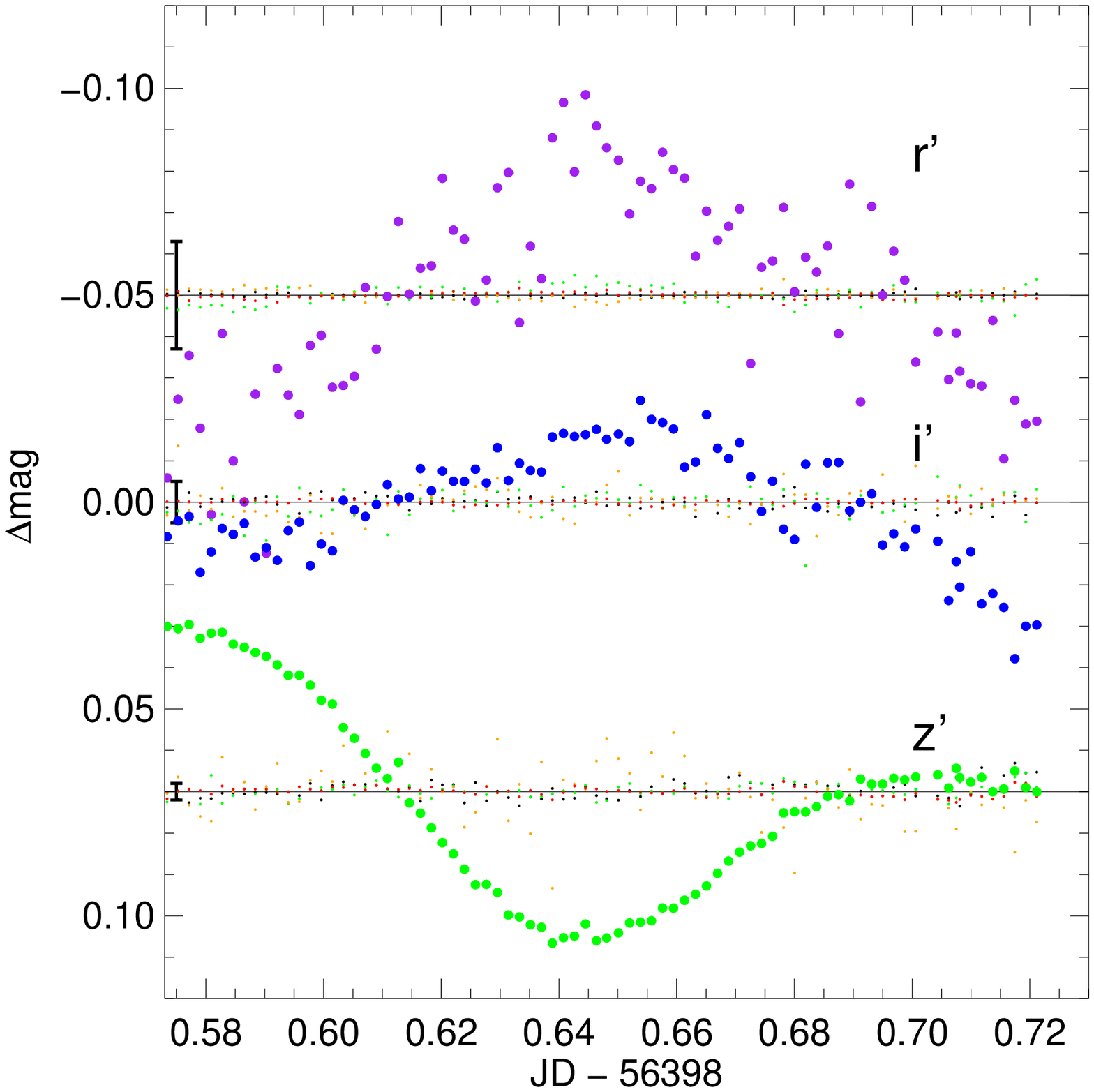}
\includegraphics[width=3.5in]{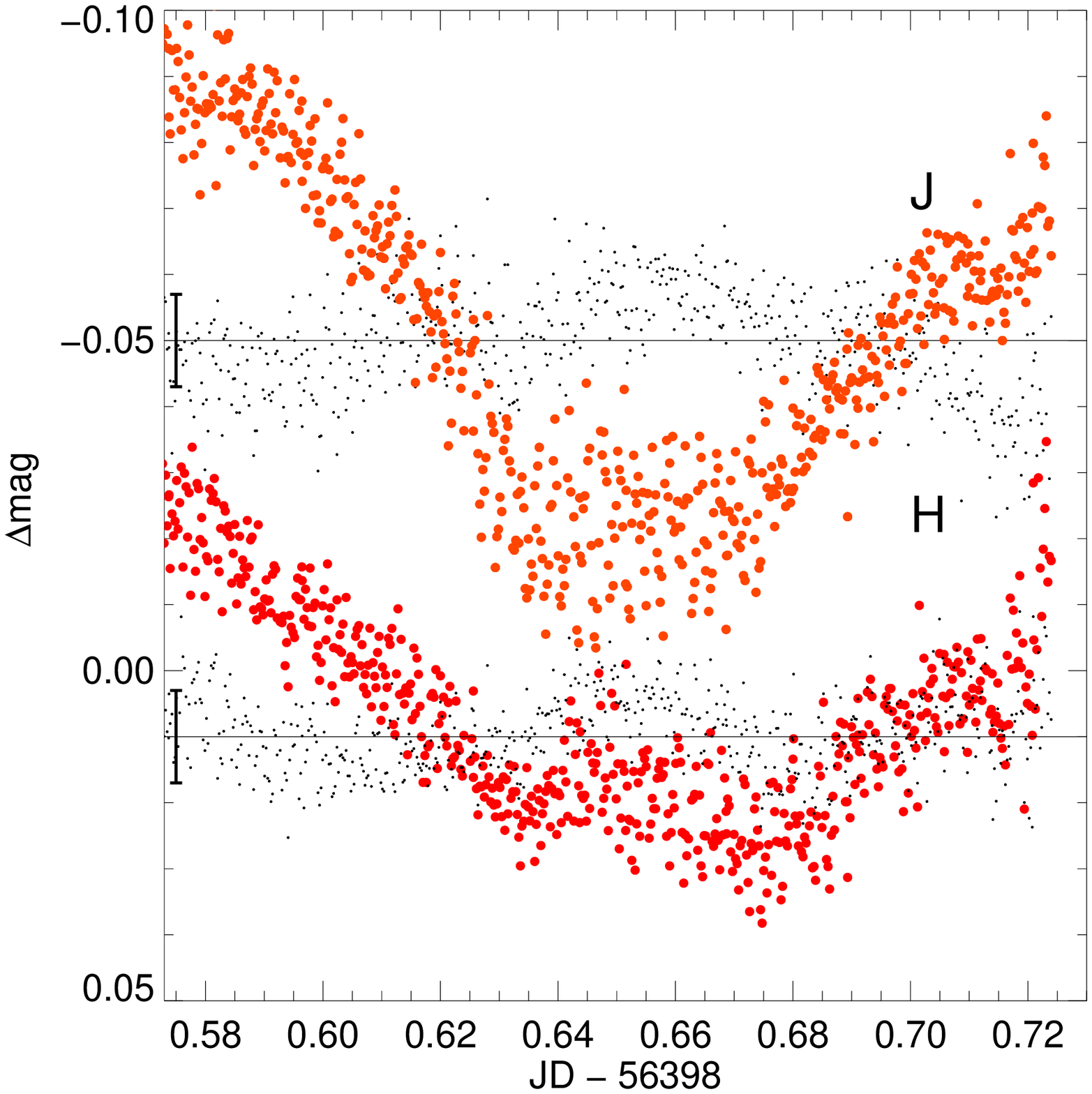}
\caption{Unresolved light curves from aperture photometry on UT 2013-04-16.
Estimated error bars
are plotted at the beginning of each light curve and 
example residual reference lightcurves (reference star - high S/N reference lightcurve)
are plotted as small dots.
\label{fig:comb_lc}
}
\end{figure}

We present optical and near-IR unresolved light 
curves from UT 2013-04-16 in Fig.~\ref{fig:comb_lc}.
For the optical data, 
the lightcurves of multiple reference stars 
(10 stars for $i'$ and $z'$, 7 stars for $r'$)  were
combined to provide a high S/N reference lightcurve
and then subtracted from the lightcurve of the brown 
dwarfs.  No significant differences were seen between the 
lightcurves of the reference stars.
We estimate our photometric uncertainty 
by subtracting a low-order polynomial from each
lightcurve to remove smooth trends (both intrinsic 
variability and instrumental effects)
and then calculating the standard deviation
of the residuals (which generally follow a gaussian 
distribution).  We find photometric uncertainties of 
13 mmag in $r'$, 5 mmag in $i'$, and 2 mmag 
in $z'$.  We estimate peak-to-trough variations over the course 
of the observation of 64 mmag (6$\%$)
in $r'$, 50 mmag (5$\%$) in $i'$, 
and 73 mmag (7$\%$) in $z'$.
The peak in brightness seen at JD$\sim$56398.65 in $r'$ and $i'$ 
correlates with a dimming at $z'$.

For the near-IR data, high S/N reference lightcurves 
were generated at $J$ and $H$ by combining 8 reference stars 
at $J$ and 7 reference stars at H.  
We find photometric uncertainties of 
7 mmag in $J$ and $H$.  We find variability (peak to trough) of 70 mmag 
(7$\%$) in $J$ and 43 mmag (4$\%$) in $H$.
Both $J$ and $H$ band light curves show dimming at JD$\simeq$56398.65 
correlated with the trough seen in brightness at $z'$ and the peaks
seen at $r'$ and $i'$.

We do not report the unresolved $K$ band 
photometry here, as we found significant differences 
between the light curve shape of reference stars and a 
very strong upward trend in all the reference stars 
in the second half of the observation in $K$, which 
varied considerably from star to star.  Thus, 
we are not confident in our ability to identify true variability of the 
brown dwarfs with these data. 


\subsection{Resolved light curves}

\begin{figure}
\includegraphics[width=3.5in]{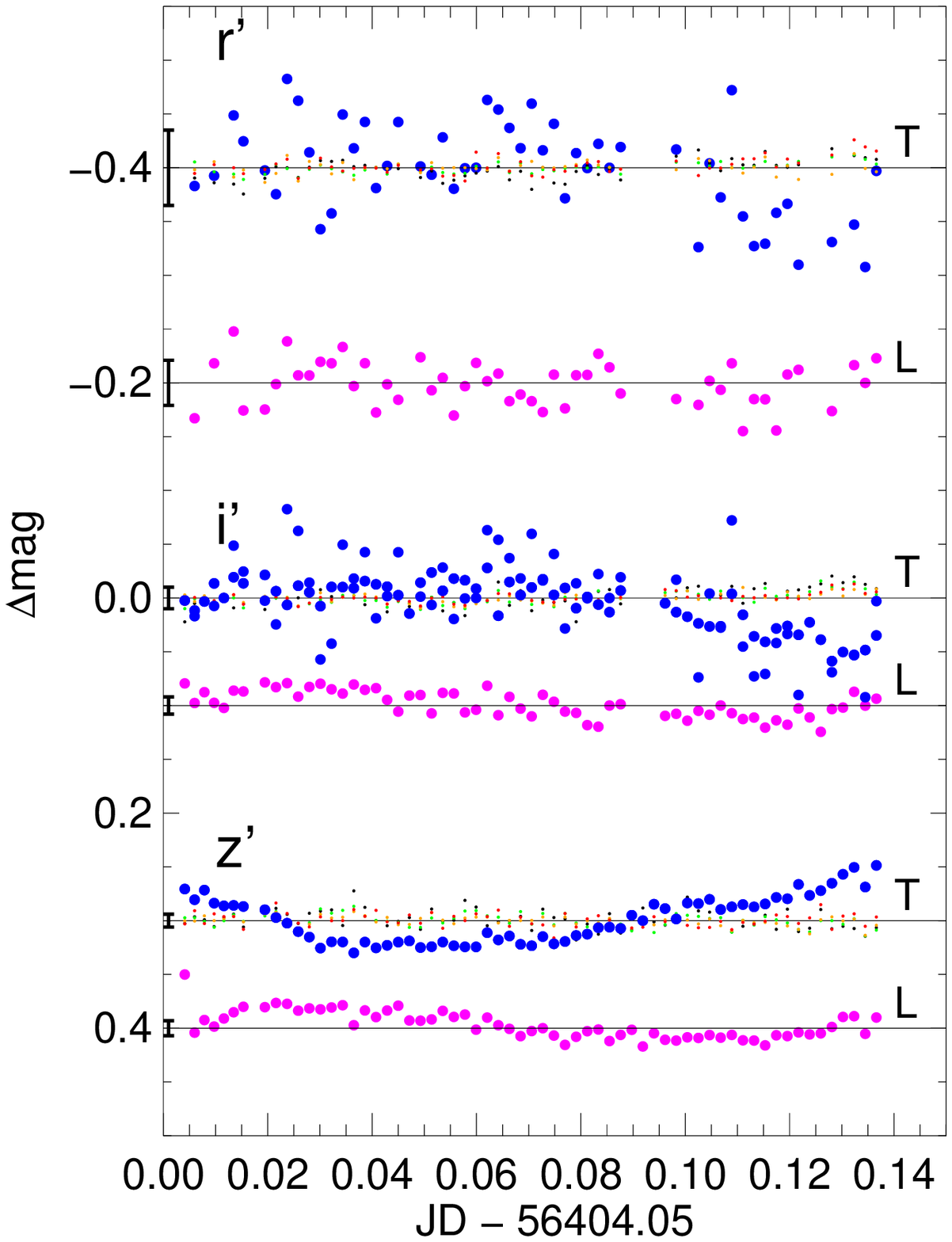}
\includegraphics[width=3.5in]{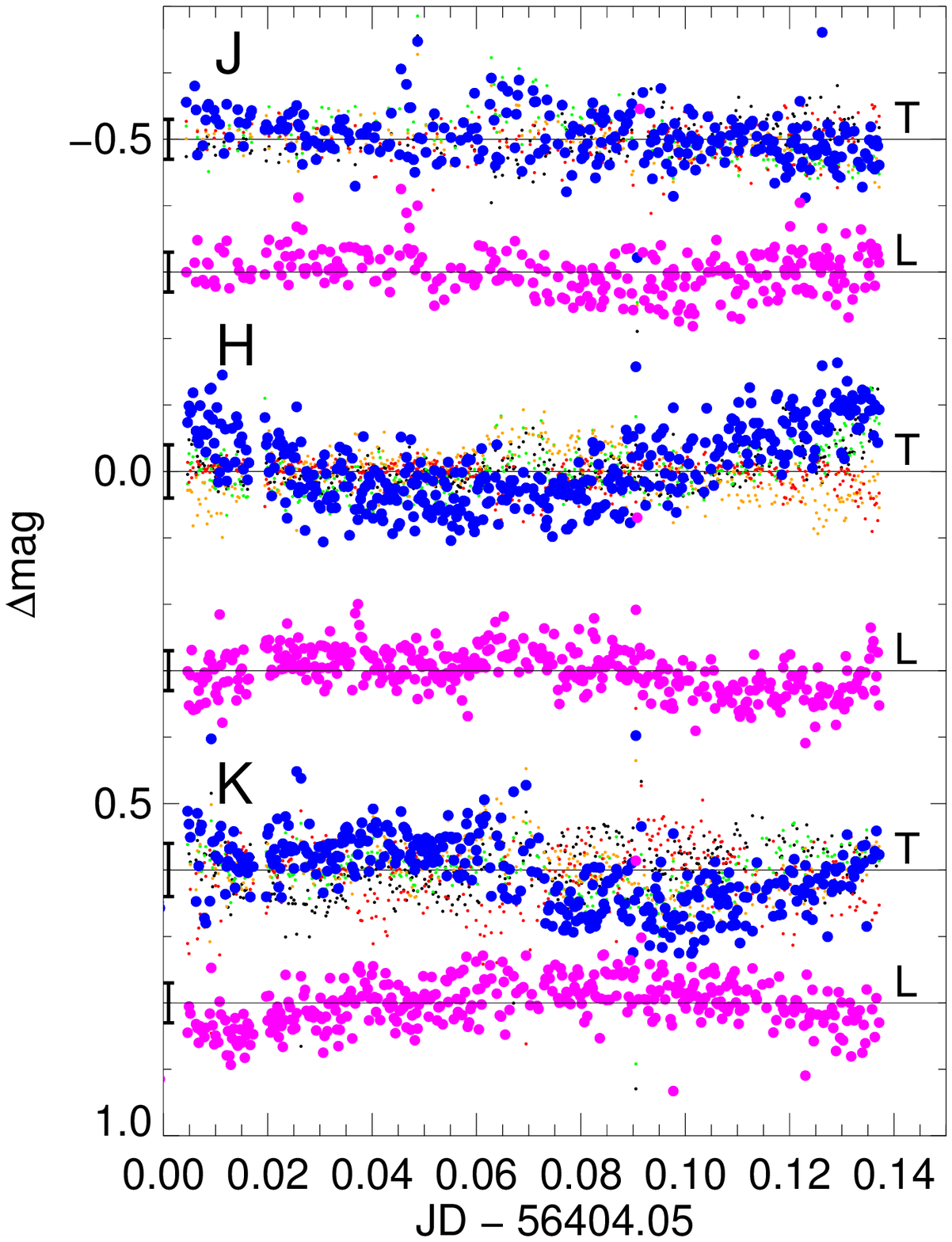}
\caption{Single component light curves from PSF-fitting photometry 
on UT 2013-04-22.  
Estimated error bars
are plotted at the beginning of each light curve
and example residual reference lightcurves (reference star - high S/N reference lightcurve)
are plotted as small dots.
\label{fig:ind_lc}
}
\end{figure}

We present optical and near-IR resolved light 
curves from UT 2013-04-22 in Fig.~\ref{fig:ind_lc}.
For the optical bands, high S/N reference lightcurves
were constructed by combining lightcurves
from 32 reference stars in $r'$, 22 reference stars in $i'$, 
and 20 reference stars in $z'$.
The binary is well-resolved and well-sampled
at optical wavelengths.
No significant differences were seen between the 
lightcurves of the reference stars.
We estimate our per-point photometric uncertainties
as described in Section~\ref{sec:unresolved}. 
For the T0.5 component, 
we measure photometric uncertainties of 
35 mmag in $r'$, 10 mmag in $i'$, and 6 mmag 
in $z'$.  We estimate variability (peak to trough) of 90 mmag 
(9$\%$) in $r'$, 60 mmag (6$\%$) 
in $i'$,  and 70 mmag (7$\%$) in $z'$.
As in the unfocused data, brightening in $r'$ and $i'$ 
appears to correlate with dimming in $z'$ (and vice versa).

\begin{figure}
\includegraphics[width=4in]{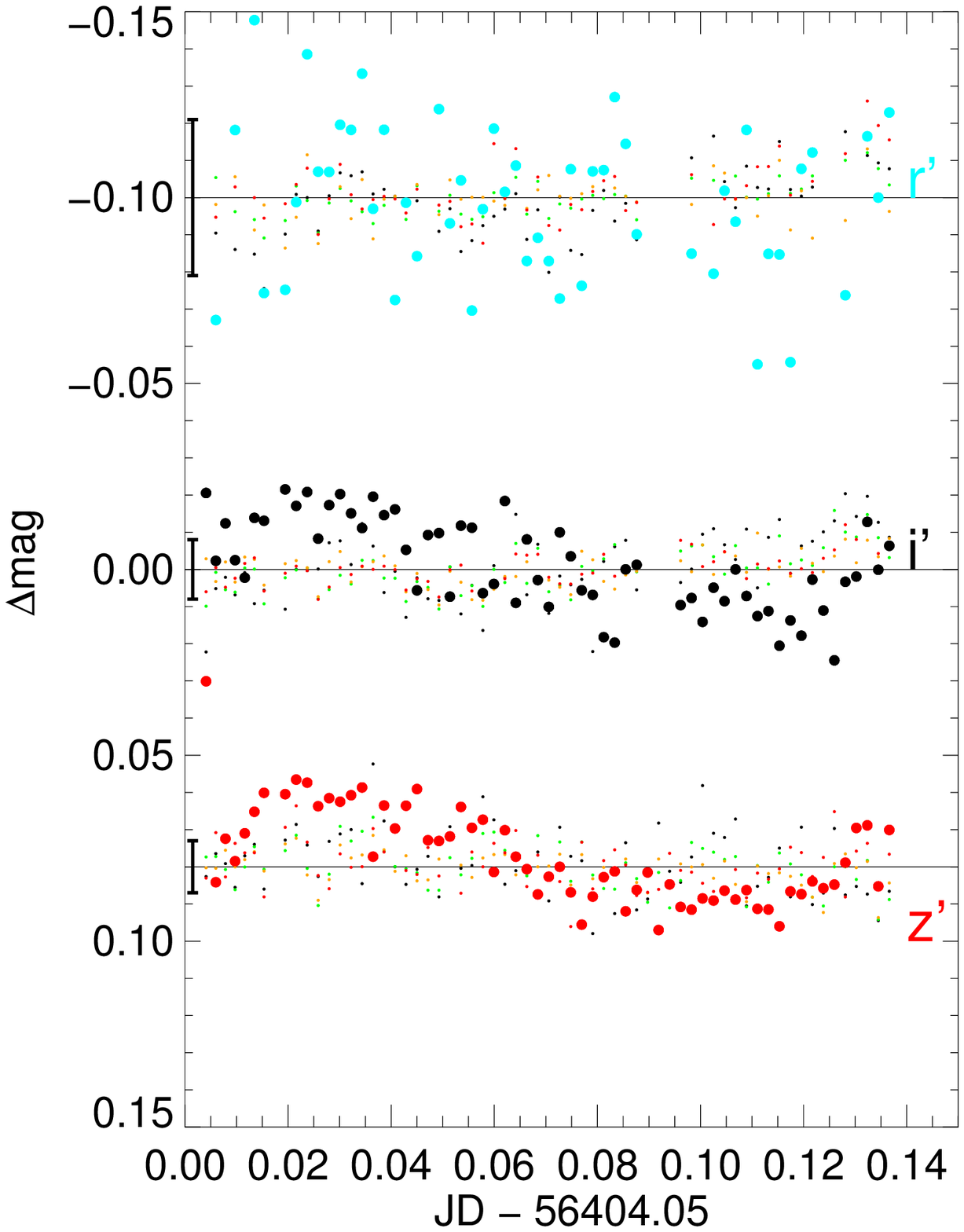}
\caption{A zoomed-in view of the optical 
lightcurve of the L7.5 component.  
\label{fig:l_opt}
}
\end{figure}

While the T0.5 component dominates the 
combined variability, the L7.5 component exhibits
variability with a smaller amplitude.   
Importantly, the L7.5 component's 
variations are uncorrelated with the variations seen in the T0.5 component. 
A zoom-in on the optical lightcurves 
of the L7.5 component is presented in 
Fig.~\ref{fig:l_opt}.  For this component, 
we find photometric uncertainties of 
21 mmag in $r'$, 8 mmag in $i'$, and 7 mmag 
in $z'$.  We find variability (peak to trough) of 
26 mmag (2$\%$) in $i'$  and 
and 29 mmag (3$\%$) in $z'$.
No variability was detected in $r'$ above the levels 
of our photometric uncertainties.
The variability appears quasi-sinusoidal at this 
epoch, with a period of 3-4 hours, but further monitoring is necessary 
to confirm this and to accurately determine 
any rotation period.

Using the Bayesian time-series analysis method of Bailer-Jones (2012),
we find that a single component sinusoidal model about the mean
which includes a variable Gaussian noise component yields a significantly higher
Bayes factor for both the $i'$ and $z'$ T0.5 component lightcurves as
well as for the $z'$ L7.5 component lightcurve than does a variable
Gaussian noise model about the mean.
However, while periodic variability has been observed for these brown
dwarfs, this periodicity is not necessarily sinusoidal. This result does nonetheless
add confidence that the observed variability is real and not a result
of noise.

For the near-IR data, high S/N reference lightcurves 
were produced at $J$, $H$, and $K$ by combining 
lightcurves from 4 stars in $J$, 5 stars in $H$, and 4
stars in $K$.  Reference stars were chosen to be similarly 
bright as the target brown dwarfs and to 
lie as close to the target brown dwarfs as possible.
For the T0.5 component, we find photometric uncertainties 
of  30 mmag in $J$ and 40 mmag in $H$ and $K$.
We measure peak to trough variability of 140 mmag (13$\%$) 
in $H$ and 110 mmag (10$\%$) in $K$.
No variability was detected in $J$ above the levels 
of our photometric uncertainty.  As with the defocused 
light curves, the minimum brightness in $H$ is close to
a similar minimum in $z'$.
The $K$ band light curve also shows a clear minimum 
brightness, offset by $\sim$0.05 days (1.2 hours) from the $H$ band minimum.  
For the L7.5 component, we estimate a photometric uncertainty
of 30 mmag in $J$, $H$, and $K$.  No variability was detected in the
near-IR for this component above the levels of our photometric 
uncertainty.

Since the L7.5 component does not vary in the near-IR within our
photometric errors, we performed a check 
of our T0.5 near-IR lightcurves by using the L7.5 lightcurve as 
our reference lightcurve.  The T0.5 component and L7.5 component lie on 
a similar part of the chip and also possess similar luminosities 
and colors, thus the L7.5 component should serve as an ideal reference lightcurve
for the T0.5 component, provided it itself is not significantly variable.  
Using the L7.5 as the reference lightcurve, we retrieved
similar relative lightcurve shapes and amplitudes in  $J$, $H$, and $K$ 
(taking into account the greater noise using a single reference)
as we did with the high S/N reference lightcurves.

\section{Discussion}

\subsection{Trends, period, and phase shifts}

We note a number of significant trends in these light curves.
In both datasets, $r'$ and $i'$ light curves appear to be anticorrelated 
with $z'$ and $H$ for the T0.5 component (which likely dominates the 
variability in the unresolved lightcurve).  In the defocused dataset, $J$ appears correlated with $z'$ 
and $H$ and anticorrelated with $r'$ and $i'$, while in the focused dataset
we measure no variability for $J$ at the level of our photometric
precision, likely due to evolution of weather phenomena between the observations.
In our focused T0.5 component lightcurves,  the $K$ band lightcurve appear to have a
significant phase offset relative to both $H$ and $z'$.  To estimate the 
magnitude of this phase offset, we must first adopt a period 
for the variability of the T0.5 component.  

Gillon et al. (2013) report a variability period of 4.87$\pm$0.01 hours 
for the system, which they attribute solely to the T0.5 component.   
We cannot confirm this period but we can confirm that the period of the T0.5
component must be greater than 3.5 hours.  
In our resolved observations, the L7.5 lightcurves
may cover a full period in $i'$ and $z'$ over the
course of 3.5 hours but this must be confirmed.
For the calculation of phase offsets we adopt the period
of 4.87$\pm$0.01 hours from Gillon et al. (2013) and assume that the effect 
of the L7.5 component on the period of the combined defocused
lightcurve is small.

For the T0.5 single component lightcurve and the combined-light
lightcurve (variability assumed to be dominated by the T0.5 component), 
we measure phase shifts between bands by fitting low-order polynomials
to our lightcurves and finding the time of minima and maxima in JD.  
Assuming that phase=0 occurs at the extremum of the $z'$ band, we calculate the 
phase shift relative to $z'$ for the other bands.
We estimate uncertainties on phase shifts of 0.005 days in the optical and 0.01 days in the near-IR.  
We assign a phase shift of 
180$^{\circ}$ for bands anticorrelated with $z'$, 
assuming periodic variability.  The
observed anti-correlation implies that the light curves are 
180$^{\circ}$ out of phase, but this phase offset must be 
confirmed in future observations covering more than one full period.

For the T0.5 component, phase shifts cluster around 0$^{\circ}$ in $z'JH$ and around
180$^{\circ}$ in $r'$$i'$, with the $K$ band lightcurve displaying a phase
shift of $\sim$100$^{\circ}$ relative to $H$ and $z'$
(Fig.~\ref{fig:pressure}, right panel).  
While most known variable brown dwarfs display correlated 
variability across different bands (e.g. Artigau et al. 2009, Radigan
et al. 2012), phase 
shifts between bands have been observed once before in the case of the 
T6.5 brown dwarf 2MASSJ22282889-431026 (Buenzli et al. 2012), 
where the phase shift was found to correlate with average model pressure probed at each 
band.

\subsection{Correlation between phase and atmospheric pressure}

To evaluate if a correlation between phase and atmospheric pressure
holds for WISE 1049B, we use the publicly available BT-Settl 1D
atmospheric models (Allard et al. 2012) 
to estimate the pressure levels probed by the GROND bandpasses.  
The BT-Settl models utilize a sophisticated cloud model which 
realistically takes into account the mixing properties of the
atmosphere, as determined by 2D radiation hydrodynamic 
simulations (Freytag et al. 2010).  We assume 
$T_\textrm{eff}=1300$~K as suggested by Burgasser et al. (2013), and 
$\log g=5.0$ as expected for an object with tens of Jupiter masses and 
an age of several-Gyr (e.g., Burrows et al. 2011).

\begin{figure}
\includegraphics[width=4in]{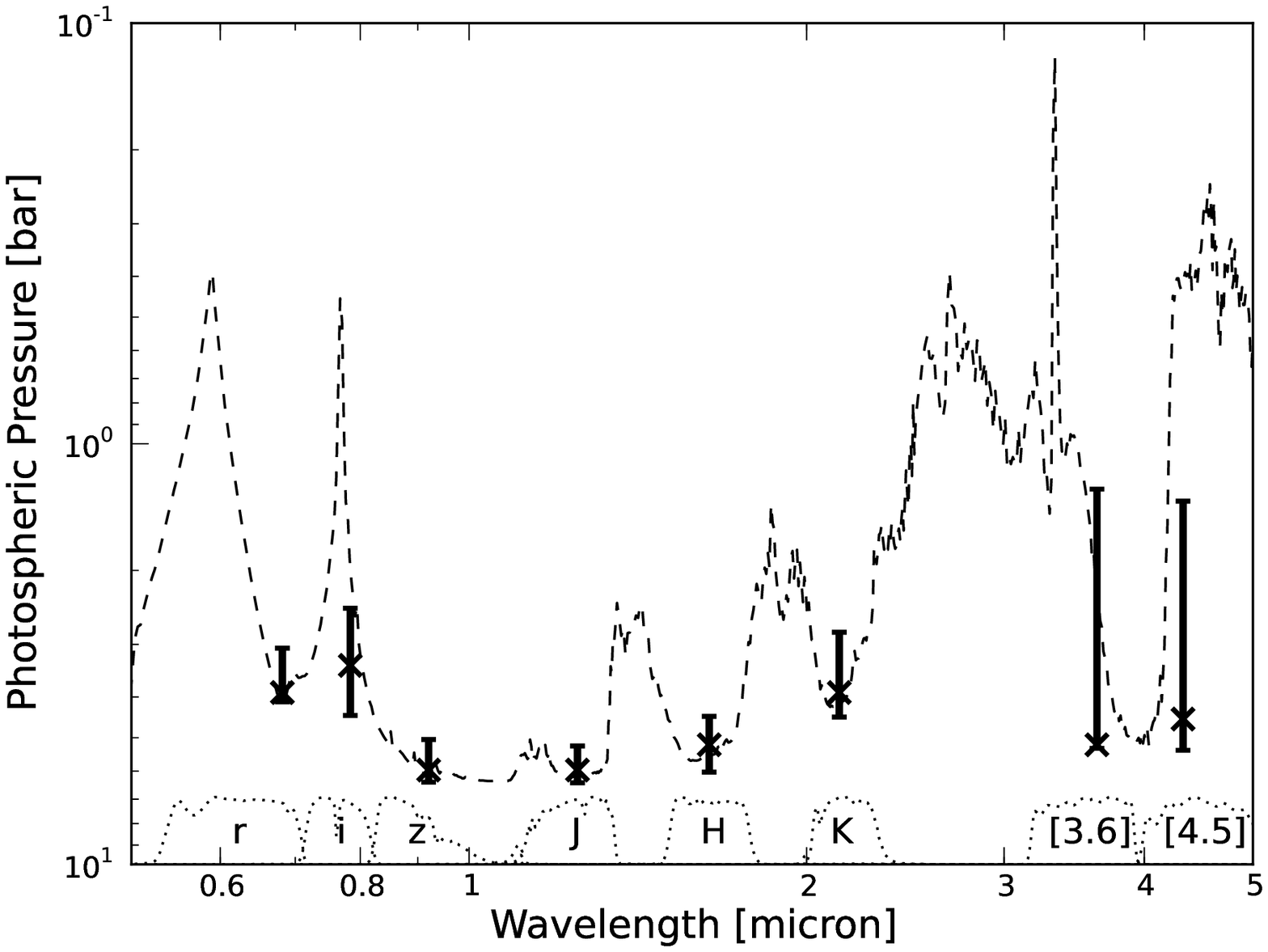}
\includegraphics[width=3in]{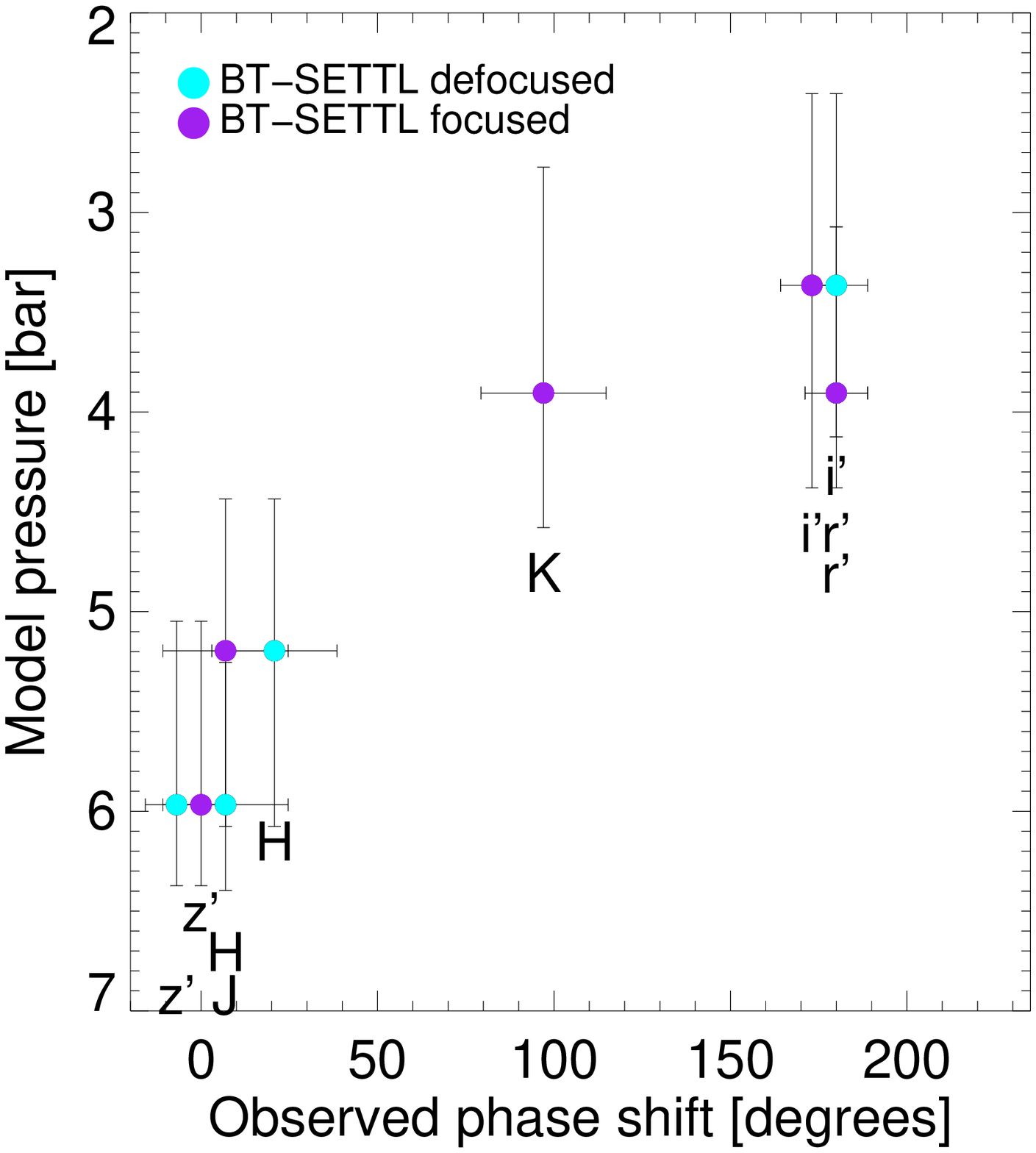}
\caption{{\bf left:} Pressure levels probed by broadband photometry of the BT-Settl 
atmospheric model described in Sec.~4.  The ``pressure spectrum'' 
 is plotted as a dashed line.  The peak of each channel's 
contribution function is marked with a ``$\times$,'' and the error bars 
indicate the pressure range from which 80\% of the flux is emitted.  
{\bf right:} Observed phase shift vs. average model atmosphere
pressure probed for both the defocused and focused observations, 
using the same atmospheric model as the left panel.
We find a correlation of
phase shift with model pressures, with flux from the higher layers
of the atmosphere phase shifted relative to the $z'$ band extremum.
\label{fig:pressure}
}
\end{figure}

We convert the model surface fluxes into brightness temperatures $T_B$. 
We then assign an effective photospheric pressure level to each 
wavelength using the model's temperature-pressure profile, which 
provides a one-to-one correspondence between $T$ and $P$. 
The resulting ``pressure spectrum'' (Fig.~\ref{fig:pressure}) shows that stronger 
absorption features correspond to deeper photospheric pressure levels. 
Using the model's emergent flux density and the product of the telluric 
and GROND filter transmission profiles, we calculate a mean photospheric 
pressure for each band and the upper and lower pressure limits within 
which 80\% of that bandpass' flux is emitted.

The bandpasses probing deepest in the atmosphere 
are $z'$, $J$, and $H$ ($P=5-6$~bar); the extremum in WISE 1049B's 
photometry occurs at nearly the same time (i.e., rotational phase) in 
these three filters, as they probe roughly the same depth 
in the atmosphere.  We also note 
that the three filters which probe higher layers of the atmosphere 
($P=3-4$~bar; in $r'$ and 
$i'$ because of absorption by alkalis, and in $K$ because of increasing molecular 
absorption) all exhibit flux extrema offset from $z'$
by $\geq$100~deg.  As plotted in Fig.~\ref{fig:pressure}, 
this result suggests that WISE 1049B exhibits a correlation 
between atmospheric depth and phase offset similar to that recently seen 
in the cooler T6.5 brown dwarf 2MASS J22282889-431026 (Buenzli et al. 
2012).  

Unlike brown dwarfs such as 2MASS J 2139 and SIMP 0136 which display
correlated variability across all wavelengths and can be modeled
as a combination of two 1D model atmospheres (a combination of
thick and patchy clouds, Apai et al. 2013), phase shifts indicate that
the vertical and horizontal structure of the atmosphere must also explicitly be
taken into account.  In particular, 1D models cannot consider
the strong effects of rotation on atmospheric structure. Showman and
Kaspi (2013) find evidence for large scale ($>$1000 km) atmospheric
circulation in three-dimensional, global, anelastic numerical
simulations of convection in rapidly rotating brown dwarfs ($<$9 hour
period).  This produces horizontal temperature variations of up to 50
K as well as significant vertical mixing, driving cloud patterns 
to evolve on 10-100 rotation period timescales.  2D radiation hydrodynamics
simulations by Freytag et al. (2010) show that gravity
waves can also play a significant role in forming and dispersing dust
clouds in brown dwarf atmospheres, but on somewhat smaller scales.  In
the case of 2M 2228, Buenzli et al. 2012 suggest a possible ``stacked cell''
scenario to produce the observed phase shifts, where two convective cells are
stacked vertically, one moving upward (ascent) the other downward
(subsidence).  Subsidence produces heating and inhibits cloud
formation while ascent has the opposite effect, cooling and
encouraging cloud formation (Buenzli et al. 2012, Showman and Kaspi
2013).  A model with two sets of stacked cells, each subtending half of the
brown dwarf could then possibly produce phase shifts.  In the case of
WISE 1049B, we suspect a similar mechanism.  The anti-correlated
lightcurves may be produced by the combination of patchy cloud layers
deeper in the photosphere (driving variability in $z'$, $J$, and $H$) and
horizontal temperature differences in the higher layers of the
photosphere (producing the $r'$, $i'$, and $K$ variability).  These features
are perhaps connected by vertical convective motions, i.e the stacked
cells suggested by Buenzli et al. (2012), explaining the
anticorrelation between $r'$, $i'$, and $z'$, $J$, $H$.  Testing this hypothesis
will require 1) data at higher spectral resolution -- our broadband
data still averages over a wide range of pressures and 2) the
development of sophisticated 3D brown dwarf circulation models coupled
with radiative transfer and cloud formation prescriptions.

\section{Conclusions}

We present two epochs of simultaneous multi-wavelength 
variability monitoring of the L/T transition brown dwarf binary 
WISE 1049AB, spanning from the optical 
through the near-IR.  We note a number of trends in our light curves.
The $r'$ and $i'$ light curves appear to be anticorrelated 
with $z'$ and $H$ for the T0.5 component and in the 
unfocused lightcurve (dominated by the T0.5).  
In the defocused dataset, $J$ appears correlated with $z'$ 
and $H$ and anticorrelated with $r'$ and $i'$, while in the focused dataset
we measure no variability for $J$ at the level of our photometric precision.
In our focused T0.5 component lightcurve, the $K$ band lightcurves appear to have a
significant phase offset relative to both $H$ and $z'$.  
We also report for the first time low-amplitude 
variability in $i'$ and $z'$ intrinsic to the L7.5 component.  We find 
that the measured phase offsets are correlated with atmospheric 
pressure probed at each band, as derived from 1D BT-Settl
atmospheric models appropriate for this object.
However, interpreting these results with 1D models, whether cloudy or clear, 
is not appropriate, as these models do not take into account 
rotation or other multi-dimensional effects which drive the atmospheric
properties of these objects.  The multidimensional models necessary to fully 
interpret these data are currently under development -- these data will
be invaluable for testing and validating future and ongoing 
multi-dimensional modeling of L/T transition brown dwarfs (Allard et
al. in prep, Showman and Kaspi 2013).



\acknowledgements
Part of the funding for GROND (both hardware as well as personnel) 
    was generously granted from the Leibniz-Prize to Prof. G. Hasinger 
    (DFG grant HA 1850/28-1).  EB is supported by the
Swiss National Science Foundation (SNSF).    The research leading to 
these results has received funding from the French
“Agence Nationale de la Recherche” (ANR), the “Programme National de Physique
Stellaire” (PNPS) of CNRS (INSU), and the European Research Council under the
European Community’s Seventh Framework Programme (FP7/2007-2013 
Grant Agreement no. 247060). It was also conducted within the Lyon 
Institute of Origins under grant ANR-10-LABX-66. 
D.H. acknowledges support from the European Research Council under the European
Community’s 
Seventh Framework Programme (FP7/2007-2013 Grant Agreement no. 247060).
Atmosphere models have been computed on the {\sl P\^ole Scientifique de
Mod\'elisation Num\'erique} (PSMN) at the {\sl \'Ecole Normale
Sup\'erieure} (ENS) in Lyon and at the {\sl Gesellschaft f{\"u}r
Wissenschaftliche Datenverarbeitung G{\"o}ttingen} thanks to a
collaborative agreement with the Institut f{\"u}r Astrophysik G{\"o}ttingen.






\clearpage

\end{document}